\documentclass[fleqn,usenatbib]{mnras}

\usepackage{newtxtext,newtxmath}
\usepackage[T1]{fontenc}
\usepackage{libertine} 
\usepackage{siunitx}

\DeclareRobustCommand{\VAN}[3]{#2}
\let\VANthebibliography\thebibliography
\def\thebibliography{\DeclareRobustCommand{\VAN}[3]{##3}\VANthebibliography}

\usepackage{graphicx}	% Including figure files
\usepackage{amsmath}	% Advanced maths commands
\title[KT Eri super-remnant ]{Introducing the Condor Array Telescope: IV. A possible nova super-remnant surrounding the putative recurrent nova KT Eridani}

\author[M. M. Shara et al.]{
Michael M. Shara,$^{1}$\thanks{E-mail: mshara@amnh.org (MMS)}
Kenneth M. Lanzetta,$^{2}$
James T. Garland,$^{1}$
Stefan Gromoll,$^{3}$
David Valls-Gabaud,$^{4}$
\newauthor 
Frederick M. Walter,$^{2}$
John K. Webb,$^{5}$
Alexei Kniazev,$^{6,7,10}$
Lee Townsend,$^{6,7}$
Matthew J. Darnley,$^{8}$
\newauthor
Michael Healy-Kalesh,$^{8}$
Jesus Corral-Santana,$^{9}$
and Linda Schmidtobreick,$^{9}$
\\
$^{1}$Department of Astrophysics, American Museum of Natural History, CPW \& 79th street, New York NY 10024-5192, USA\\
$^{2}$Department of Physics and Astronomy, Stony Brook University, Stony Brook, NY 11794-3800, USA\\
$^{3}$Amazon Web Services, 410 Terry Ave. N, Seattle, WA 98109, USA\\
$^{4}$Observatoire de Paris, LERMA, CNRS UMR 8112, 61 Avenue de l'Observatoire, 75014 Paris, France\\
$^{5}$Institute of Astronomy, University of Cambridge, Madingley Road, Cambridge CB3 0HA, United Kingdom\\
$^{6}$South African Astronomical Observatory, Observatory Road, Observatory 7925, Capetown, South Africa\\
$^{7}$Southern African Large Telescope, PO Box 9, Observatory 7935, South Africa.\\
$^{8}$Astrophysics Research Institute, Liverpool John Moores University, Liverpool L3 5RF, UK\\
$^{9}$European Southern Observatory, Alonso de C\'{o}rdova 3107, Vitacura, Casilla 19001, Santiago de Chile, Chile\\
$^{10}$Special Astrophysical Observatory, Nizhnij Arkhyz, Karachai-Circassia, 369167, Russia}

\date{Accepted XXX. Received YYY; in original form ZZZ}

\pubyear{2023}

\begin{document}
\label{firstpage}
\pagerange{\pageref{firstpage}--\pageref{lastpage}}
\maketitle

\begin{abstract}
Just 10 recurrent novae (RNe) - which erupt repeatedly on timescales shorter than one century - are known in our Galaxy. The most extreme RN known (located in the Andromeda galaxy), M31N 2008-12a, undergoes a nova eruption every year, and is surrounded by a vast nova ``super-remnant'', 134 pc in extent. Simulations predict that {\it all} RNe should be surrounded by similar vast shells, but previous searches have failed to detect them. KT Eri has recently been suggested to be a RN, and we have used the Condor Array Telescope to image its environs through multiple narrowband filters. We report the existence of a large ($\sim$ 50 pc diameter), H$\,\alpha$-bright shell centered on KT Eri, exactly as predicted. This strongly supports the claim that KT Eri is the 11th Galactic recurrent nova, and only the second nova known to be surrounded by a super-remnant. SALT spectra of the super-remnant demonstrate that its velocity width is consistent with that of M31-2008-12a.  

\end{abstract}


%%%%%%%%%%%%%%%%%%%%%%%%%%%%%%%%%%%%%%%%%%%%%%%%%%

%%%%%%%%%%%%%%%%% BODY OF PAPER %%%%%%%%%%%%%%%%%%

\section{Introduction}
\subsection{Novae}
Nova eruptions occur near the surfaces of white dwarfs (WDs) accreting matter from close binary companions \citep{Warner1995}. As hydrogen-rich matter accumulates in an envelope on a WD, degenerate electron pressure increasingly dominates gas pressure at the envelope base. Because the pressure of degenerate electrons is temperature independent, accreted envelopes do not initially expand as nuclear reactions and their consequent heating get underway. As a result, powerful thermonuclear runaways (TNR) develop wherein most CNO nuclei undergo proton-capture in just a few minutes. The result, a classical nova eruption \citep{Starrfield1972,Prialnik1978}, converts several percent of the accreted hydrogen to helium, unbinds the WD envelope and achieves peak luminosities up to $10^{5} - 10^{6}$\,L$_\odot$ for days to months. Mass ejection ends, and a nova eruption ``shuts off'' when the accreted envelope is ejected \citep{Prialnik1979}. Once accretion resumes the WD's envelope grows until it again erupts as a nova. All novae must erupt thousands of times over the course of their multi-Gyr lifetimes \citep{Ford1978}.

\subsection{Recurrent novae}
The time between successive nova eruptions depends strongly on the mass of the accreting WD. Suites of nova models \citep{Yaron2005} demonstrated that low mass (0.6 to 0.8 M$_\odot$) WDs with relatively large radii and weak surface gravities must accumulate massive envelopes ($10^{-3} - 10^{-4}$\,M$_\odot$) to generate enough pressure to initiate nova TNRs. This can take a Myr or more, so that nova eruptions on low mass WDs are expected to be relatively rare, as is observed \citep{Shara2018}. More massive WDs ($\sim$ 0.8 - 1.1 M$_\odot$) with higher surface gravities can accumulate critical-mass envelopes which yield TNRs in just thousands of years \citep{Yaron2005}. Novae containing these more massive WDs are the ones seen to erupt most frequently in any given few centuries \citep{Shara2018}, the time-span in which humans have been detecting and recording nova eruptions. WDs with masses in excess of $\sim$ 1.2 M$_\odot$ can accumulate critical-mass envelopes, and generate ``recurrent'' novae separated in time by less than one century \citep{Hillman2016}. While such massive WDs are rare, their frequent eruptions suggest that roughly a quarter of all the Galactic nova eruptions of the past century are recurrent novae \citep{Pagnotta2014}. Most have not yet been seen to erupt a second time, but they will be as high-cadence digital all-sky surveys (like the Rubin Observatory) locate and classify millions of transients every night. 

The time between successive nova eruptions also depends on the average rate of mass transfer $dM \, / \, dt$ in a pre-nova binary. Higher $dM \, / \, dt$ will more rapidly bring an accreted envelope to the critical mass (for a given WD mass) to initiate a nova TNR. The larger radii and smaller surface gravities of subgiant and giant donors enable them to transfer mass faster than main sequence stars. It is thus not surprising that nearly-Chandrasekhar mass WDs with post main sequence companions are the ideal recipe for making recurrent novae. 

\citet{Pagnotta2014} pointed out that recurrent novae tend to display modest outburst amplitudes, long orbital periods, and red infrared colors (all due to their post main sequence donors). In addition, they exhibit high mass ejection velocities, high excitation spectral lines, an eruption light curve with a plateau, and a WD mass greater than 1.2 M$_\odot$ (all due to their massive WDs). They then singled out several likely recurrent novae masquerading as ``ordinary'' novae, including KT Eri (see section 1.5). 

The most rapidly recurring nova known (M31N 2008-12a, hereafter simply ``12a") is situated in the Andromeda galaxy. Its annual nova eruptions \citep{Darnley2014} must be due to a nearly Chandrasekhar-mass WD \citep{Hillman2016}, which needs to accrete just $\sim 10^{-7}$\,M$_\odot$ to trigger a TNR. The donor is either a low luminosity red giant or a horizontal branch (red clump) star \citep{Darnley2017} which is capable of transferring mass to its WD at a rate $dM \, / \, dt$ that is high enough ($\sim$ $10^{-7}$\,M$_\odot$/yr) to enable annual nova eruptions. The powerful TNRs on the WD of 12a lead to mass ejection near the beginning of the outburst with velocities as high as 13,000\,km\,s$^{-1}$ \citep{Darnley2016}, which then decrease throughout an outburst \citep{Darnley2017}. 

\subsection{Nova super-remnants}
An inevitable conclusion of the observed range of ejection velocities in 12a is that the early, fast ejecta of the (N+1)th eruption must overtake, collide and mix with the late, slower ejecta of its Nth eruption. In addition, the outermost ejecta must have encountered and swept up surrounding ISM, and thus been decelerated, making it a slow-moving target for later eruptions' ejecta. There is evidence for this phenomenon in the ejecta of the recurrent nova T Pyx \citep{Schaefer2010}, where previously faint knots have ``turned on". 

Groundbased and {\it Hubble Space Telescope} images have revealed a vast ring-like structure of ejecta and swept-up ISM surrounding 12a \citep{Darnley2019}. At 134 pc in size it is larger than most supernova remnants, and has been dubbed the first (and to date only known) nova ``super-remnant''.  The phenomenon of successive ejecta colliding with each other and with surrounding ISM should describe not just 12a, but {\it all} RNe \citep{Healy-Kalesh2023}. These authors' extensive simulations demonstrated that nova super-remnants encompassing RNe should be $\sim$20 - 120 parsecs in extent. If this prediction is correct, then why is the 12a super-remnant the only one known? 

\subsection{Detecting super-remnants} 
Depending on the mass ejected and the density of the surrounding ISM, the shells of novae erupting infrequently should fade on timescales of decades to multiple centuries \citep{Tappert2020} after eruption. The shells of GK Per (nova Per 1901 CE) \citep{Duerbeck1987,Shara2012b} and Nova Sco 1437 CE \citep{Shara2017b} remains easily detectable with 1-m-class telescopes, while the much fainter 2000+ year old ejecta of Z Cam is only partially revealed in multi-hour exposures with 4-m class telescopes \citep{Shara2007,Shara2012a}. Much deeper imaging, designed to reach the lowest surface brightnesses ever achieved for nova ejecta, have ``filled in'' the previously ``missing'' sections of Z Cam's shell, and hint at a {\it second} larger, concentric and still-fainter shell \citep{Shara2023} over 1 degree in diameter. We posit that {\it much more sensitive and larger area imaging} than has yet been accomplished is essential to detecting most hitherto unseen nova ejecta, including nova super-remnants, due to their extremely low surface brightnesses. A first test of this hypothesis is the essence of this paper.

\subsection{KT Eridani - a recurrent nova?}

KT Eridani (Nova Eridani 2009) was a fast nova (fading from peak brightness m$\sim$5.4 by 3.0 mag in 13.6 d), whose light curve was serendipitously captured by the Solar Mass Ejection Imager (SMEI) instrument on the {\it Coriolis} satellite in exquisite detail \citep{Hounsell2010}. \citet{Pagnotta2014}'s suggestion that KT Eri might be a recurrent nova was strengthened by the detailed analysis of \citet{Schaefer2022}. They determined an orbital period of 2.61595 d and a companion star temperature of 6200 $\pm$ 500 K, hence a subgiant companion; an absolute system V magnitude of + 0.7 $\pm$ 0.3, implying a very high $dM \,/\, dt$ of 3.5 x $10^{-7}$\,M$_\odot$/yr; light and radial velocity curves that demonstrate a white dwarf mass of 1.25 $\pm$ 0.03 M$_\odot$; and a likely recurrence timescale $\sim$ 50 yr (though only one eruption has been recorded to date). 

That a nova which reaches naked-eye brightness at maximum light could have been missed in the 1950s or 1960s might seem surprising. But as emphasized by \citet{Pagnotta2014}...``Eruptions over the last century can easily be missed if they
happen when the star is too close to the Sun, during the full moon, during any of many intervals
when no one was watching, or if the search did not go deep enough even if someone was searching in the right area at the right time. For CNe, the discovery efficiency is 22\% for novae peaking at V = 6 mag and 9\% for novae peaking at V = 10 mag, even in ideal conditions; this rate remains fairly constant from 1890 to 2012." That previous eruptions of KT Eri, early in and during the middle of the 20th century, have likely been missed is not surprising at all.  

At a {\it Gaia}-determined distance \citep{Bailer-Jones2021} of $\sim$ 5 kpc, the predicted $\sim$ 20-120 pc diameter shell surrounding KT Eri should subtend an angle of 0.22 - 1.3 degrees. This is larger than the fields of view of almost all previous nova shell searches (e.g. \citealt{Sahman2022}), so it is not surprising that no shell associated with KT Eri is known. Thus motivated, we undertook a program of narrow-band imaging of KT Eri as part of the science verification of the new Condor Array Telescope \citep{Lanzetta2023b,Lanzetta2023a}.

In Section 2 we describe the capabilities of Condor and the imaging of KT Eri with it. In Section 3 we present the Condor images which display the large, low surface brightness H$\,\alpha$ shell surrounding KT Eri. The spectra and the measured angular and spatial sizes of the shell are discussed in section 4, and our results are briefly summarized in Section 5.

\section{Observations}

\subsection{The Condor Array Telescope}
\label{sec:condor} % used for referring to this section from elsewhere
Condor consists of six apochromatic refracting telescopes of objective diameter 180 mm, each equipped with a large-format (9576 × 6388 pixels), very low read-noise (1.2 e$^{-1}$), very rapid read-time (< 1 s) CMOS camera. Though the six telescopes are aligned on the same point in the sky, they have no interferometric capability. In addition to Sloan g', r' and i' filters, this ``array'' telescope is equipped with a set of narrow-band filters (each of 3 nm FWHM, one per telescope). The ions (central wavelengths) corresponding to those six filters are  $\ion{He}{II}$ (468.6~nm), $[\ion{O}{III}]$ (500.7 nm), $\ion{He}{I}$ (587.6 nm), H$\,\alpha$ (656.3 nm), $[\ion{N}{II}]$ (658.4 nm), and $[\ion{S}{II}]$ (671.6 nm). Condor is located at the Dark Skies New Mexico observatory near Animas, New Mexico. A full description of the telescope is given in \citet{Lanzetta2023b,Lanzetta2023a}.

\subsection{Condor Data}
\label{sec:data} % used for referring to this section from elsewhere
We used Condor with its complement of narrow-band filters to observe KT Eri between October 2021 and November 2022. The Condor CMOS cameras obtain an image scale of 0.85 arcsec per pixel, which results in a 134 × 91 arcmin$^2$ field of view. Each exposure was 600 s in length. The observation log, filter central wavelengths and the ``reach" = the net observation time multiplied by the aperture area through each filter (a useful metric for comparing array telescopes) are summarised in Table~\ref{tab:observations}. The array was dithered by a random offset of 15 arcmin between exposures. Images of the dusk and dawn twilight sky were obtained every night, and bias observations were obtained at the end of every night.

The observations were processed through the Condor data pipeline (Papers I and II), with steps involving bias subtraction, flat fielding and background subtraction, astrometric calibration, and photometric calibration. As described in \citet{Lanzetta2023b}, the astrometric calibration yields systematic uncertainties of $\lesssim$ 0.1 arcsec. Difference images were constructed using the saccadic fast Fourier transform (SFFT), an image subtraction algorithm and software package introduced in \citet{Hu2022}.

\begin{table}
\centering
\caption{Details of observations of Condor field 044754-101043 which surrounds KT Eri, centered at RA = 04:47:54, Dec = $-$10:10:43 (J2000).}
\label{tab:observations}
\begin{tabular}{lccr}
\hline
\multicolumn{1}{c}{Filter} & \multicolumn{1}{c}{Start} & \multicolumn{1}{c}{End} & \multicolumn{1}{c}{Reach} \\
\multicolumn{1}{c}{} & \multicolumn{1}{c}{Date} & \multicolumn{1}{c}{Date} & \multicolumn{1}{c}{m$^2$ s} \\
\hline
 $\ion{He}{II}$ 468.6 nm      & 2021-10-26 & 2022-11-27 & 6,382.1 \\
 $[\ion{O}{III}]$ 500.7 nm    & 2021-10-26 & 2022-11-27 & 6,183.6 \\
 $\ion{He}{I}$ 587.5 nm       & 2021-10-26 & 2022-11-27 & 6,015.6 \\
 H$\,\alpha$ 656.3 nm    & 2021-10-26 & 2022-11-27 & 5,893.5 \\
 $[\ion{N}{II}]$ 658.4 nm     & 2021-10-26 & 2022-11-27 & 5,542.3 \\
 $[\ion{S}{II}]$ 671.6 nm     & 2021-10-27 & 2022-11-27 & 3,389.5 \\
\hline
\end{tabular}
\end{table}

\subsection{SALT data}

Two spectra of the KT Eri shell, each one hour in length, were obtained with the Southern African Large Telescope (SALT) \citep{O'Donoghue2006} and its Robert Stobie Spectrograph \citep{Burgh2003,Kobulnicky2003} using its PG2300 grating on 22 and 23 February 2023. The grating/spectrograph covered the wavelength range 6510 - 6725 {\AA} with a resolution of 2.2 $\pm$ 0.25 {\AA}. The data were reduced with the SALT RSS pipeline described in \citet{Kniazev2022}, which includes correction of bad columns and lines on the CCDs; calculation and application of gain-correction coefficients; construction of a spectral flat field for subsequent pixel sensitivity correction; and location and removal of cosmic ray events. These steps are followed by reduction of a reference spectrum; reduction of a spectrophotometric standard, and reduction of the object. 

\begin{figure*}
    \includegraphics[width=0.9\textwidth]{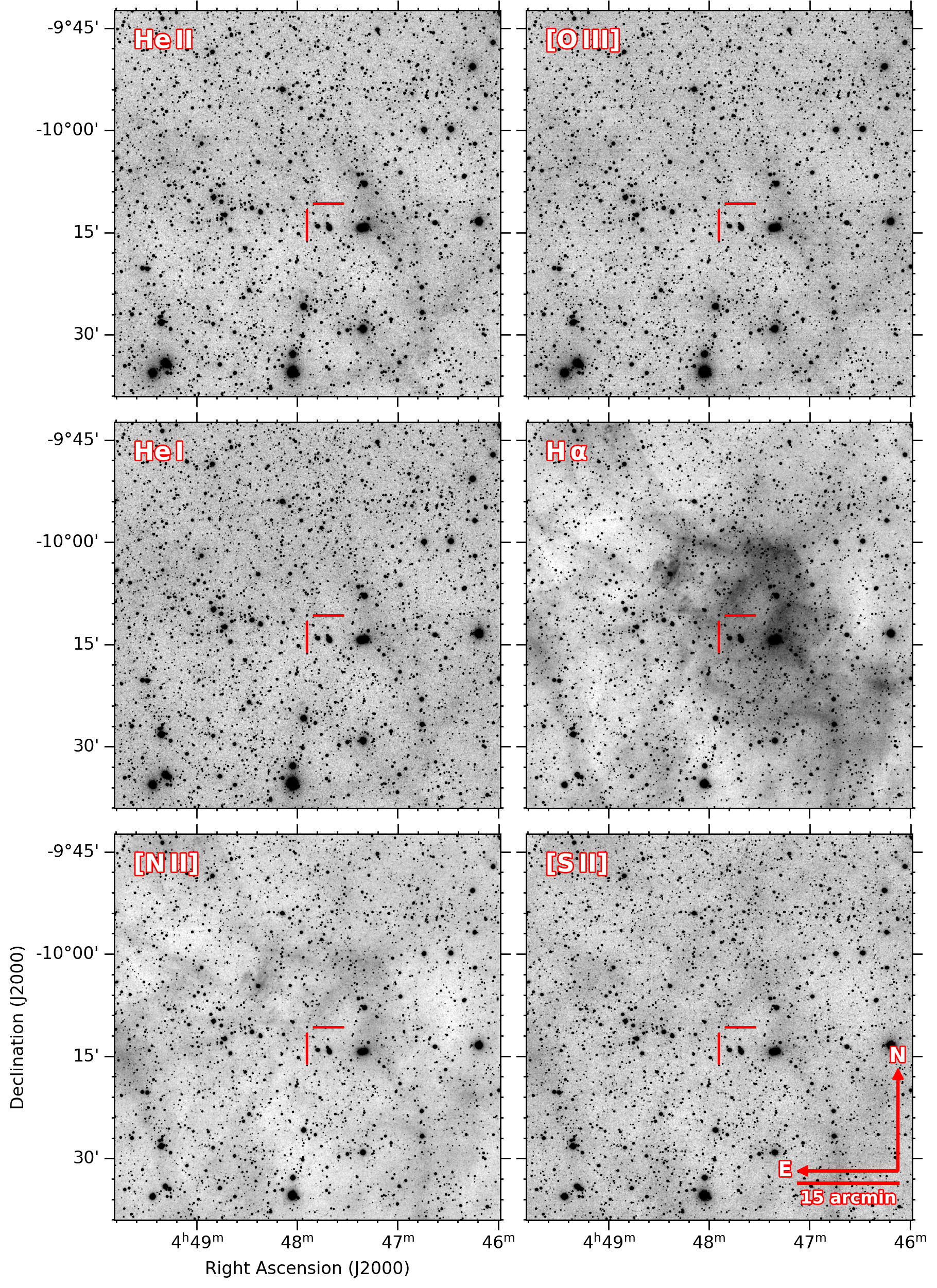}
    \caption{Clockwise from the top right, Condor images of the area surrounding KT Eri in $[\ion{O}{III}]$, H$\,\alpha$, $[\ion{S}{II}]$, $[\ion{N}{II}]$, $\ion{He}{I}$, and $\ion{He}{II}$. KT Eri is marked with red ticks. The total exposure time in each of the six narrow-band filters multiplied by the aperture for that filter (the ``reach'') is listed in Table~\ref{tab:observations}. The total exposure time in the H$\,\alpha$ filter was 14,475 sec. Each image was taken with one of the six 180 mm apochromatic refracting telescopes that constitute the Condor Array \citep{Lanzetta2023b}. Images have been smoothed via a Gaussian kernel ($\sigma = 3.0$ pixels) and are displayed with linear scaling.}
    \label{fig:six_panel}
\end{figure*}

\begin{figure*}
    \includegraphics[width=0.9\textwidth]{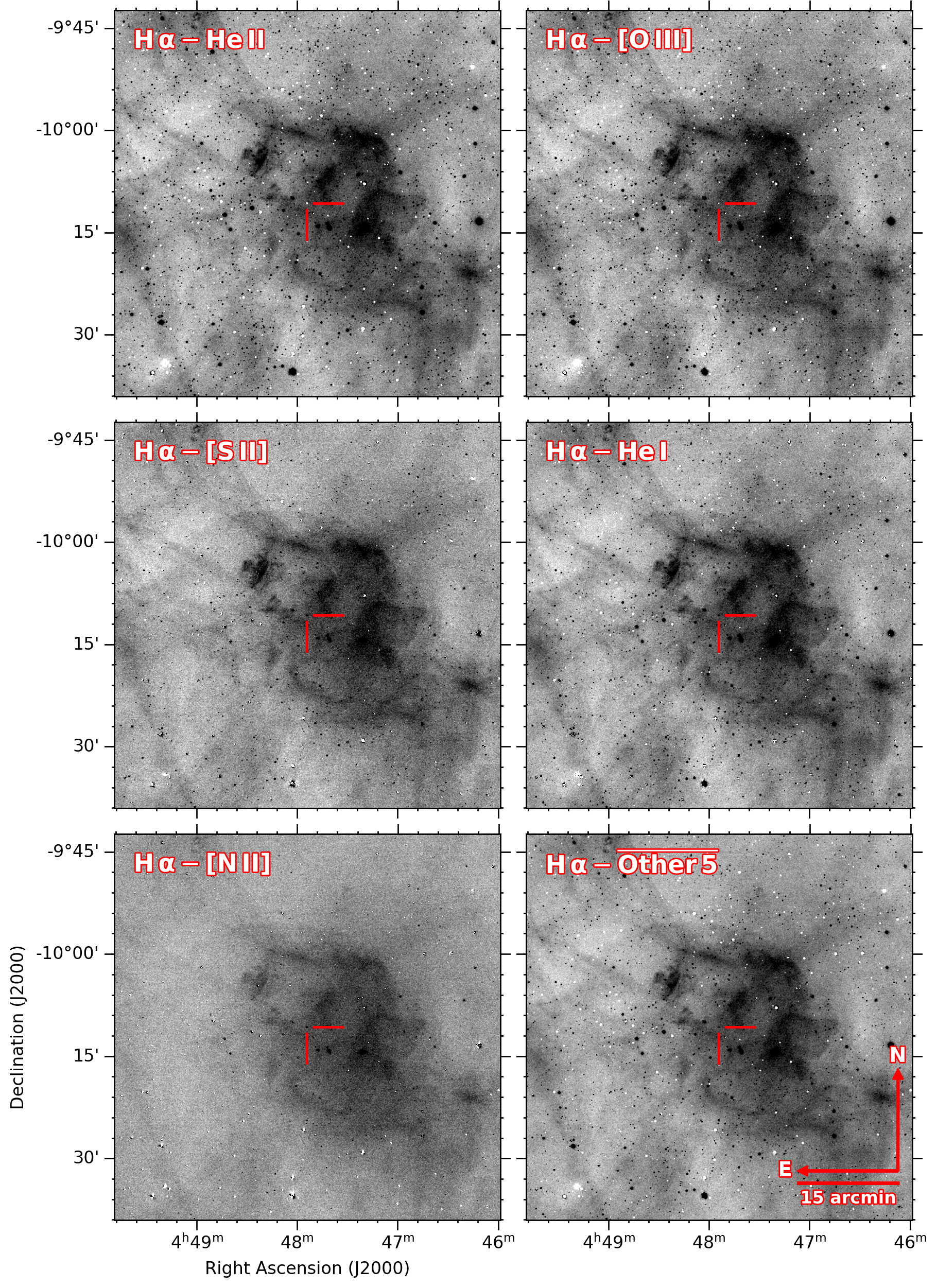}
    \caption{Clockwise from the top right, difference images of the area surrounding KT Eri between H$\,\alpha$ and $[\ion{O}{III}]$, $\ion{He}{I}$, the mean of the other five bands, $[\ion{N}{II}]$, $[\ion{S}{II}]$, and $\ion{He}{II}$. KT Eri is marked with red ticks. Images have been smoothed via a Gaussian kernel ($\sigma = 2.5$ pixels) and are displayed with linear scaling.}
    \label{fig:six_panel_sub}
\end{figure*}

\begin{figure*}
    \includegraphics[width=0.8\textwidth]{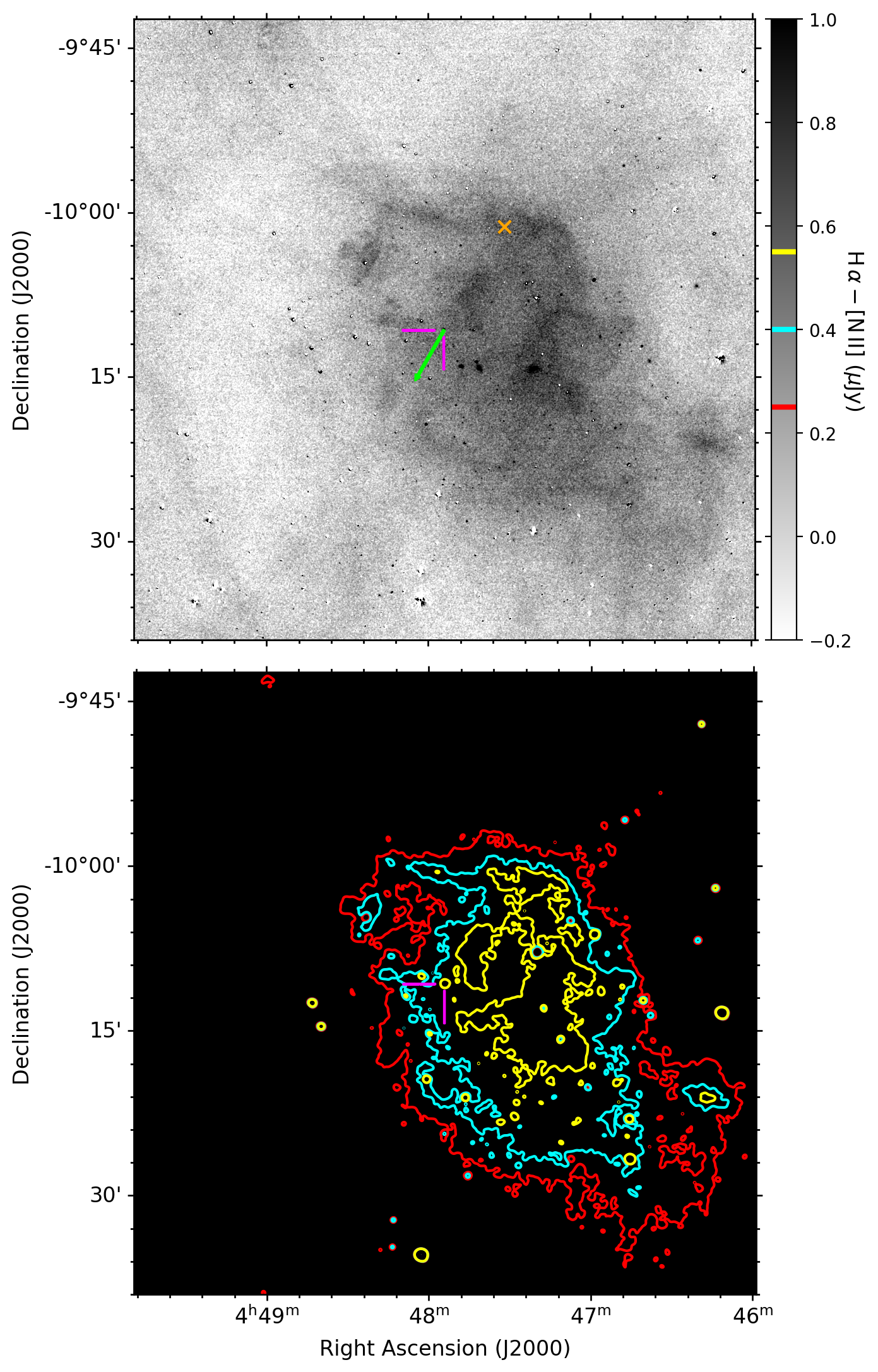}
    \caption{H$\,\alpha - [\ion{N}{II}]$ difference image (top) and associated contours (bottom) of the region surrounding KT Eri. KT Eri is marked in both images with magenta ticks. The proper motion of KT Eri, projected forward 25,000 years, is shown with a green arrow. Also shown with an orange $\times$ is the position of KT Eri 51 kyr ago. As denoted in the color bar next to the top frame, red, cyan, and yellow contours are drawn at flux density differences of 0.25, 0.40, and 0.55 $\mu$Jy, respectively. The image has been smoothed via a Gaussian kernel ($\sigma = 2.5$ pixels) and is displayed with linear scaling.}
    \label{fig:Ha_NII_sub}
\end{figure*}

\begin{figure*}
    \includegraphics[width=0.95\textwidth]{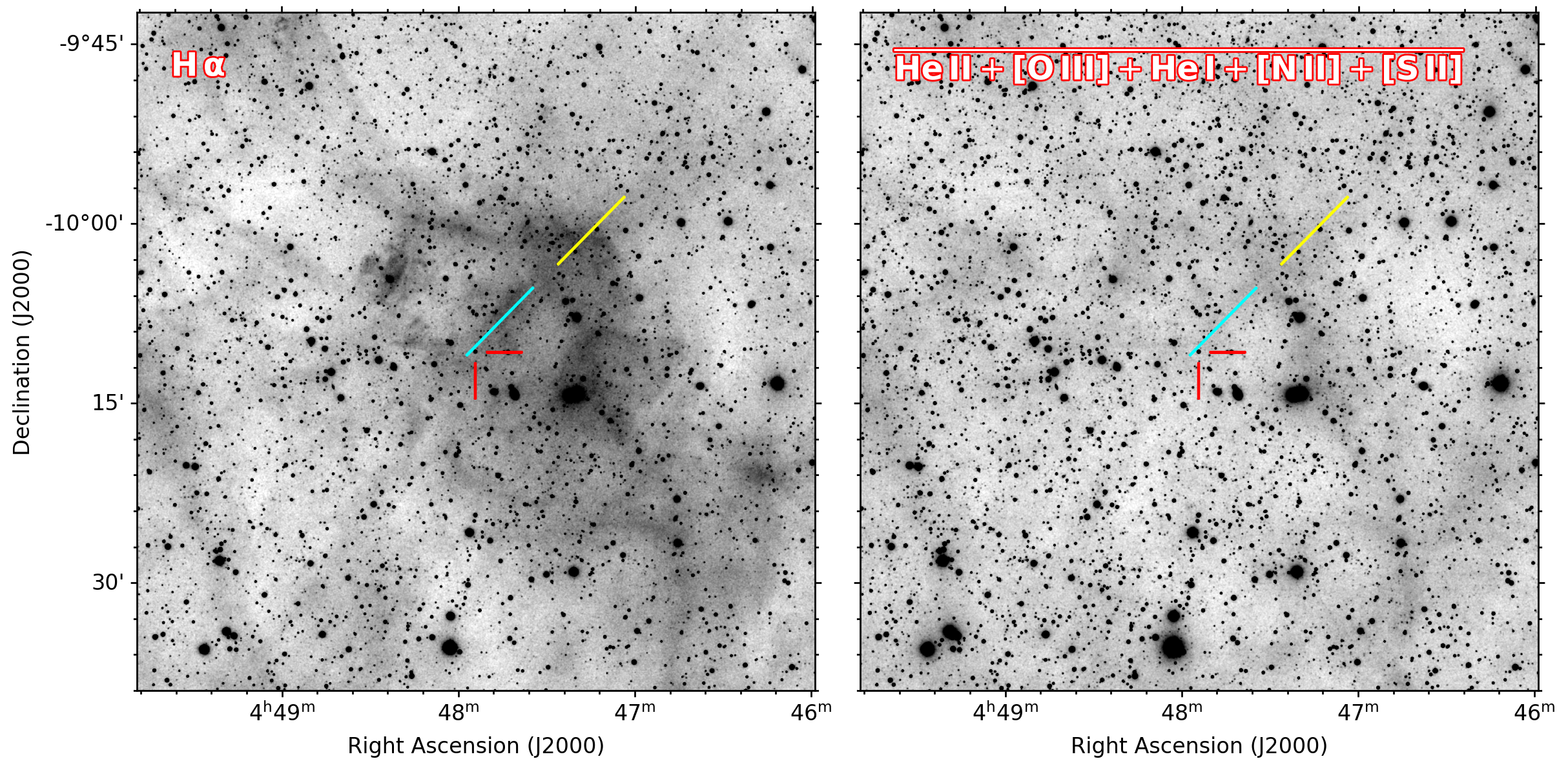}
    \caption{H$\,\alpha$ image (left) and averaged $[\ion{O}{III}]$, $[\ion{S}{II}]$, $[\ion{N}{II}]$, $\ion{He}{I}$, and $\ion{He}{II}$ (right) of the region surrounding KT Eri. KT Eri is marked in both images with red ticks. The positions of the inner and outer SALT RSS slits are shown as yellow and cyan lines, respectively (see Figs.~\ref{fig:PG2300_inside},~\ref{fig:PG2300_outside}). The images have been smoothed via a Gaussian kernel ($\sigma = 3$ pixels) and are displayed with linear scaling.}
    \label{fig:Ha_all_slits}
\end{figure*}

\section{The environment, extended nebulosity and age of KT Eri}

 \subsection{Environment}
The Galactic coordinates of KT Eri are ($207.9863^{\circ}, -32.0202^{\circ}$). At a \textit{Gaia} distance of $5110^{+920}_{-430}$ pc from adopting a Bayesian prior on the stellar density along the line of sight (see \citealt{Schaefer2022}), KT Eri lies $\sim$ 3 kpc below the Galactic Plane. Its radial velocity relative to the Sun is $-142 \pm 5$\,km\,s$^{-1}$ \citep{Schaefer2022}. 

KT Eri lies in the direction of the Orion-Eridanus superbubble \citep{Heiles1998,Boumis2001,Frisch2007}, but far beyond it.   The superbubble is a set of overlapping supernova remnants which may be associated with the Orion OB1 association, centered $\sim$ 800 pc from the Sun. Superbubbles and their associated supernovae give rise to Galactic chimneys \citep{Norman1989} and hot gas in the Galactic halo \citep{Henley2010}. \citet{Hill2012} modeled the time-dependent, supernova-driven, turbulent, magnetized Interstellar Medium below the Galactic Plane. They found that warm (T $\sim 10^{5}$ K) and hot gas (T $> 10^{6} $ K) with densities of $\sim 10^{-3} - 10^{-4}$\,cm$^{-3}$ alternately dominate at |z| $\sim$ 3 kpc, and that large-scale vertical oscillations of the chimney gas are ongoing. Differential rotation of the Galactic disk, and gas heating mechanisms in addition to supernovae \citep{Henley2015} further complicate any effort to predict the velocity of gas, relative to the Sun, in the neighborhood of KT Eri. 

\subsection{Nebulosity}
The six Condor narrow-band images of KT Eri are shown as a mosaic in Fig.~\ref{fig:six_panel}, with KT Eri indicated by a pair of red tick marks. The coherent, shell-like structure surrounding KT Eri is strongest by far in H$\,\alpha$, where the typical surface brightness is $4\times10^{-17}$$\mathrm{erg} \, \mathrm{s}^{-1} \, \mathrm{cm}^{-2} \, \mathrm{arcsec}^{-2}$.

Much fainter patterns of nebulosities are seen in each of the three images of $\ion{He}{II}$, $[\ion{O}{III}]$ and $\ion{He}{I}$.  The large range of ionization potentials of these three ions suggests that we are not detecting their emission lines in our narrowband images, but rather continuum light scattered by Galactic cirrus. There are much weaker indications of $[\ion{N}{II}]$ and $[\ion{S}{II}]$ filter-image nebulosities that mimic the pattern seen in H$\,\alpha$, so we cannot rule out that these are line-emitting regions. But the very weak [SII] argues against the nebulosity in Figure 1 being due to a supernova remnant.

The five difference images (H$\,\alpha$ minus the other five filters), as well as H$\,\alpha$ minus the (appropriately scaled) sum of the other five narrowband images, are shown in Fig.~\ref{fig:six_panel_sub}. It is again apparent that faint nebulosity pervades the entire region, and that the $\sim$ 0.6 deg diameter structure surrounding KT Eri is (by far) most apparent in H$\,\alpha$. 

In the top panel of Fig.~\ref{fig:Ha_NII_sub} we again show the H$\,\alpha - [\ion{N}{II}]$ image. The two filters' wavelengths are separated by just 2.1 nm , giving the best subtraction of stars and continuum light possible. The {\it Gaia} proper-motion vector is also shown, indicating KT Eri's motion towards the SE. The lower panel of Fig.~\ref{fig:Ha_NII_sub} shows three brightness contours (red, cyan and yellow contours corresponding to H$\,\alpha - [\ion{N}{II}]$ flux density differences of 0.25, 0.40 and 0.55 $\mu$Jy, respectively). 

\subsection{Motion of KT Eri and the shell's age}

The angular size of the shell, if at the $\sim$ 5 kpc distance of KT Eri, corresponds to $\sim$ 50 pc. This is comparable (logarithmically) in size to the 134 pc super-remnant of 12a, but two orders of magnitude larger than the $\sim$ 0.1 - 1 pc-sized ejecta of 19th and 20th century novae. KT Eri last erupted as a nova in 2009, and the speed of its ejecta was $>$ 4000\,km\,s$^{-1}$ \citep{Yamaoka2009}. Even ignoring deceleration as those ejecta interact with the surrounding interstellar medium \citep{Duerbeck1987}, it will take the ejecta $\sim$ 6000 yr to achieve the observed size of the KT Eri shell. 

As shown in the accompanying paper (Healy-Kalesh et al. 2023, in press) which models the interactions of KT Eri's successive ejecta with each other, the currently observed shell can be grown to its observed size in no less than 25 - 51 kyr. Given the \textit{Gaia}-derived parameters of KT Eri, we are able to contextualize its position and motion relative to the Earth and Galactic Center over the timescale of the shell's formation. In particular, we can pose and answer the question: was KT Eri inside its nebulosity 50 kyr ago, as must have been the case if the nova is responsible for the nebulosity now surrounding it? Using the GALA galactic dynamics package \citep{Gala2017,Price-Whelan2020} and a multi-component Milky Way potential model with disk parameters from \citet{Bovy2015}, we integrated the orbit of KT Eri backwards 51 kyr from \textit{Gaia}'s 2016.0 reference epoch. Over this time, KT Eri moved 14.3 pc with respect to the Galactic Center, or approximately 0.29 shell diameters. 

To determine the sensitivity of the computed displacement of KT Eri to the adopted orbital parameters, we again adopted the Bovy model for the Galactic potential. We then computed 800 random orbits for KT Eri which sampled the multidimensional Probability Distribution Function assuming Gaussian errors in proper motions, including the covariance in RA and DEC, radial velocity and distance. For the distance to KT Eri we used a uniform distribution between 5110+920 pc and 5110-430 pc. Looking back 51 kyr gives a median distance traveled by the binary of 18-20 pc, somewhat larger than above, but still well within the bounds of the observed shell. This result demonstrates that 1) the likely uncertainty in how far KT Eri has traveled since its putative shell began forming is just $\sim$ 3-5 pc.

In summary, KT Eri's continued presence within its own shell is consistent with the shell's ejection history and the binary's motion within the Galaxy.

\section{SALT spectra of the shell}

\begin{figure*}
    \includegraphics[width=0.9\textwidth]{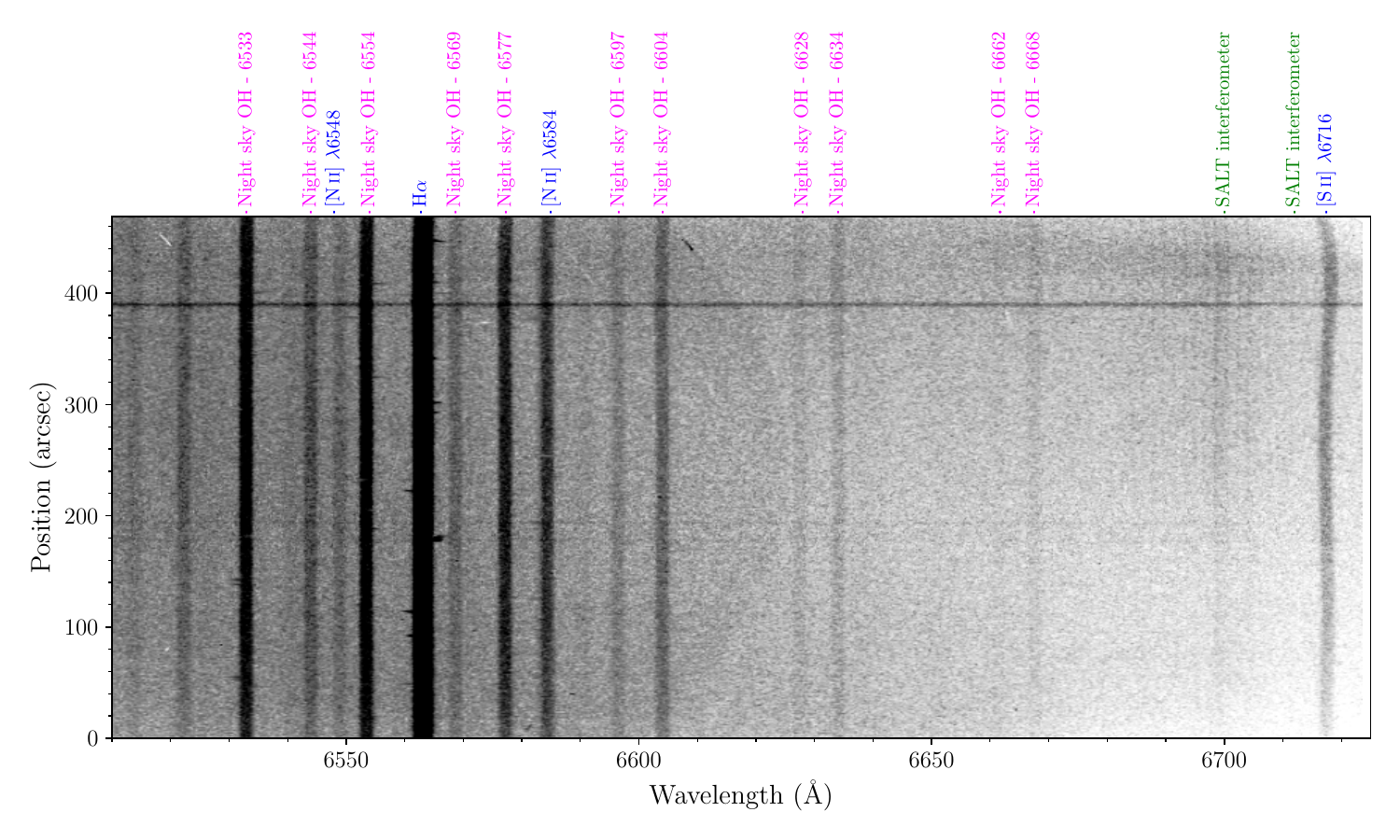}
    \caption{SALT spectrum of the inner KT Eri shell. A few sources are background galaxies. A few short horizontal line segments (especially on H$\,\alpha$) are artifacts due to imperfectly removed cosmic rays. All faint unidentified lines are due to the night sky. }
    \label{fig:PG2300_inside}
\end{figure*}

\begin{figure*}
    \includegraphics[width=0.9\textwidth]{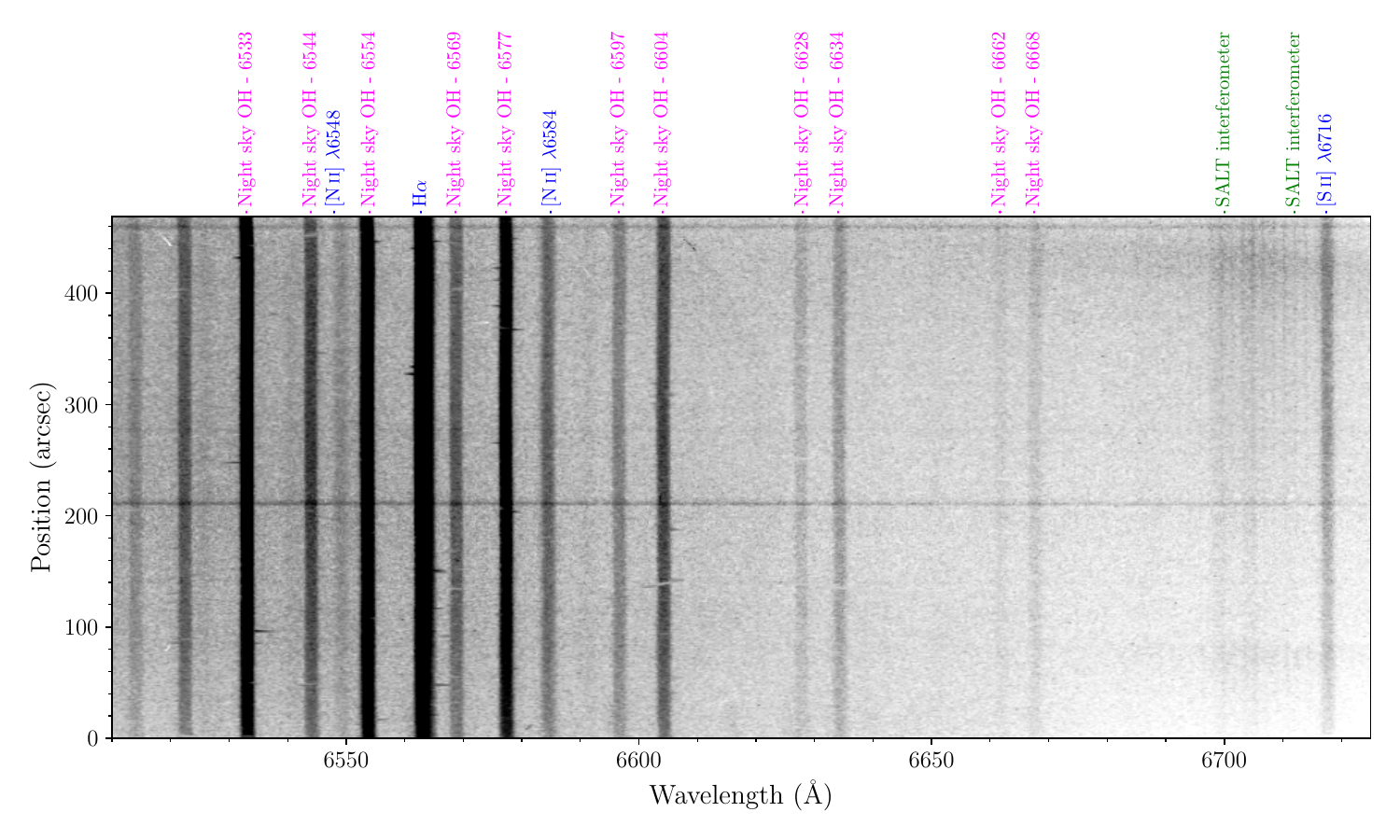}
    \caption{SALT spectrum of the outer KT Eri shell. A few sources are background galaxies. A few short horizontal line segments (especially on H$\,\alpha$) are artifacts due to imperfectly removed cosmic rays. All faint unidentified lines are due to the night sky.}
    \label{fig:PG2300_outside}
\end{figure*}

The locations of the two slit positions used in our RSS observations are shown in Fig.~\ref{fig:Ha_all_slits}, and the reduced 2-D spectra are shown in Figs.~\ref{fig:PG2300_inside} and \ref{fig:PG2300_outside}. The faint, curved features between $\sim$ 660.0 - 662.5 nm in the the 2D spectra are residual artifacts due to slightly imperfect correction of the gain of the two amplifiers on each of the RSS CCD detectors. The slight ``wiggle" in the [SII] line at 671.6 nm in Fig.~\ref{fig:PG2300_inside} is also an artifact. The few point sources are background galaxies.

\begin{figure*}
    \includegraphics[width=0.85\textwidth]{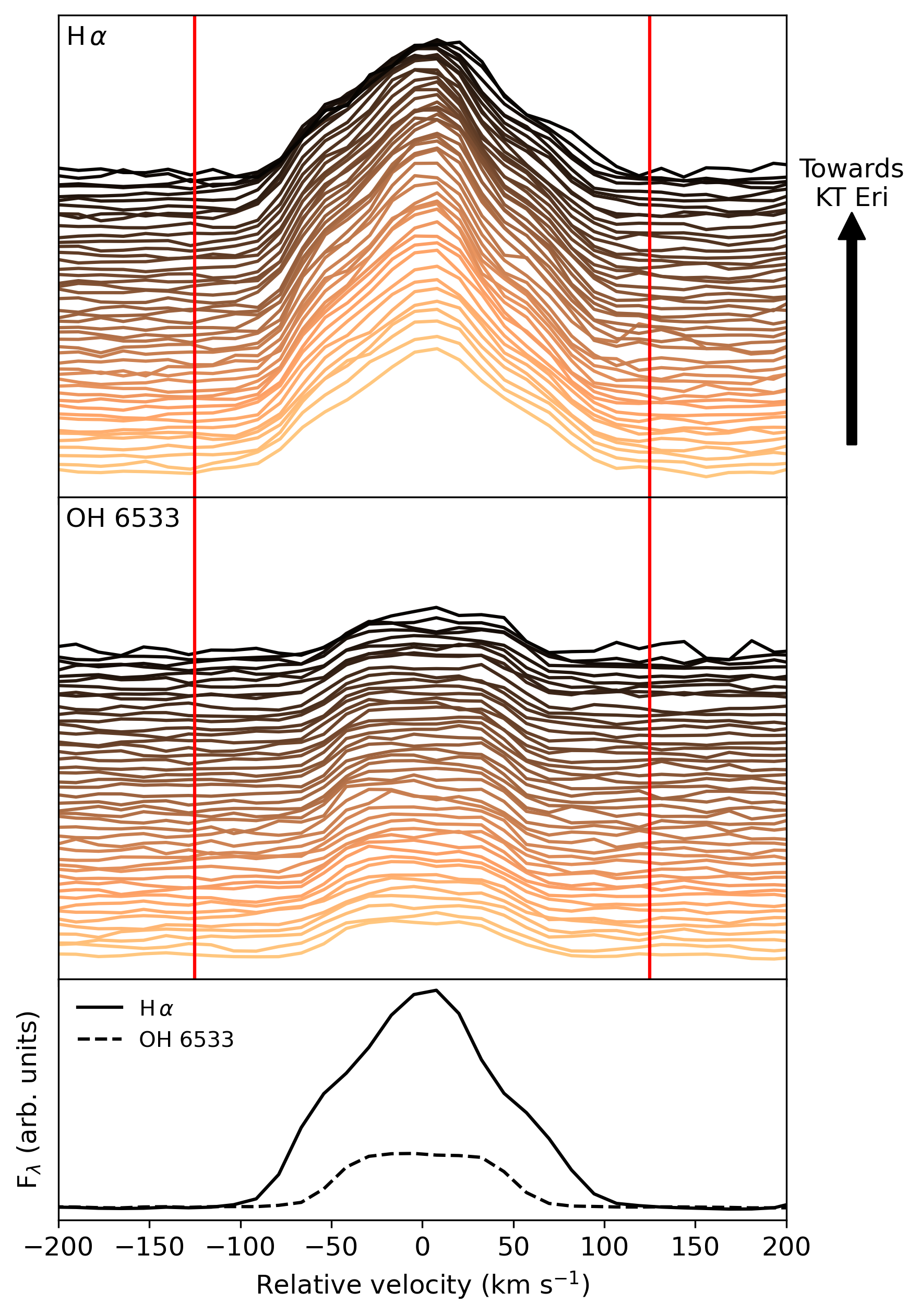}
    \caption{Spatially-binned SALT H$\,\alpha$ and OH 653.30 nm spectra of the inner shell slit. The spectra are displayed relative to the lines' respective rest velocities. In the top two panels, each profile is given by the median of 10.4 arcsec bins along the length of the slit. The binned spectra are colored and displaced to clarify the evolution of the spectra over the slit, becoming darker and moving upwards in the direction of KT Eri. Vertical red lines at $\pm125$\,km s$^{1}$ correspond to the approximately .55 nm full width at zero intensity of the H$\,\alpha$ line. The median-binned profiles along the entire length of the slit are shown in the bottom panel, with H$\,\alpha$ and OH 653.30 nm as solid and dashed lines, respectively.}
    \label{fig:binned_inside}
\end{figure*}

\begin{figure*}
    \includegraphics[width=0.85\textwidth]{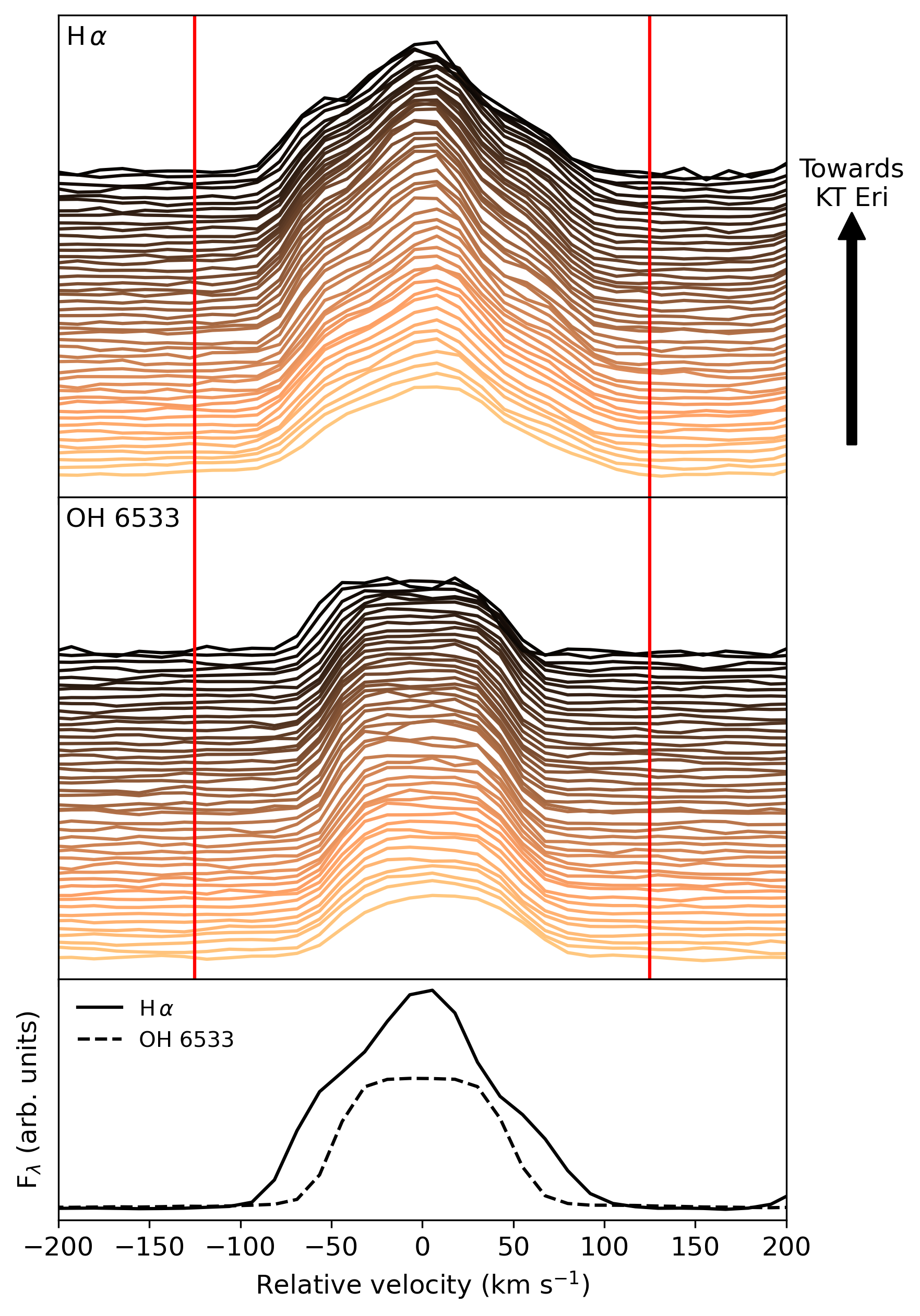}
    \caption{The same as Fig.~\ref{fig:binned_inside} for the outer shell slit.}
    \label{fig:binned_outside}
\end{figure*}

The profiles of the $H\,\alpha$ line in the inner and outer part of the shell, compared to the nearest night-sky OH line, are shown in Figs.~\ref{fig:binned_inside} and \ref{fig:binned_outside}, respectively. The H$\,\alpha$ line is the {\it only} line resolved in the spectra. It is broader than any of the other lines, all of which are due to the night sky, including $[\ion{N}{II}]$ and all of the OH lines. H$\,\alpha$ 656.28, OH 653.30, and $[\ion{N}{II}]$ 658.35 nm have full widths at zero intensity of approximately .55, .45, and .45 nm respectively. This demonstrates that there is H$\,\alpha$ emission with velocity components displaced by up to $\pm125$\,km\,s$^{-1}$ with respect to the Sun in the KT Eri nebular spectra. There is no velocity gradient in the $H\,\alpha$ line across the nebula. The lack of higher velocities (multiple hundreds or thousands of km/s) argues (as does the lack of strong [SII] emission) against the nebulosity of Figure 1 being due to a supernova remnant.

\subsection{A simple shell model}

 The simplest shell model is spherical, uniform and uniformly-expanding. It would be expected to yield a combined spectral profile of one central peak (the bright edges of the shell) with symmetric broad wings (the rest of the shell). The centroid, spacing, and breadth of these expected features are dictated by the velocity of the shell with respect to Earth, its expansion rate and turbulent peculiar velocities within the shell's leading edges.

 At the radial velocity of KT Eri ($-142 \pm 5$\,km\,s$^{-1}$), the rest wavelength of H$\,\alpha$ emission is blue-shifted by .31 nm to 655.97 nm. If the shell mass centroid shares KT Eri's space velocity, the line-of-sight components would appear up to $\pm.55$ nm symmetrically about 655.97 nm, assuming a (model-dependent) maximum expansion velocity of $250$\,km\,s$^{-1}$ (Healy-Kalesh et al. 2023). In this scenario, these shell components would be obscured by the OH 655.36 nm and H$\,\alpha$ 656.28 nm lines in the inner slit spectrum. If the shell shares the radial velocity of KT Eri ($-142 \pm 5$\,km\,s$^{-1}$) then the brighter limb component at the edge of the shell should be unaffected by nearby lines in the outer slit spectrum; it is not seen in our observations. But in fact we expect the ejecta to ``snowplough'' many multiples of its own mass over the course of $\sim$ 50 kyr, becoming turbulent, nonspherical and nonuniform in the process, and slowing to match the velocity of the local ISM.
 
 \subsection{Effects of ISM on shell radial velocity}  As noted in section 3.1, the structure and radial velocity of the chimney-driven ISM near KT Eri is both time variable and complex. We conclude that while the $\pm125$\,km\,s$^{-1}$ width of the H$\,\alpha$ line that we observe is a good match to that of the one known nova super-remnant \citep{Darnley2019}, the velocity of the ISM near KT Eri is too complex to model, so that the small radial velocity of the nebula surrounding KT Eri neither supports nor refutes its putative status as a nova super-remnant.    

%%%%%%%%%

%%%%%%%%%%%

\section{Conclusions}

On the basis of its modest outburst amplitude, long orbital period, red infrared colors, high mass ejection velocities, high excitation spectral lines, an eruption light curve with a plateau, and a white dwarf mass greater than 1.2 M$_\odot$, \citet{Pagnotta2014} and \citet{Schaefer2022} have suggested that the classical nova KT Eri is the 11th Galactic recurrent nova. Motivated by this prediction, and by the vast nova super-remnant of 12a, we have used the Condor Array Telescope to carry out deep, 6-passband narrowband imagery of the $\sim$ 3.5 deg$^2$ surrounding KT Eri. An H$\,\alpha$-bright nebulosity that is $\sim$ 0.6 deg$^2$ in extent surrounds KT Eri. If it is co-located with KT Eri then its $\sim$ 50 pc size is far larger than the remnant of any other nova except for the annually recurring nova 12a - it is the second discovered nova super-remnant. The shell should have an age of $\sim$ 50 kyr, and KT Eri's observed space velocity is consistent with the nova being inside the shell. SALT spectra of the nebula show hydrogen with velocities up to $\pm125$\,km\,s$^{-1}$, consistent with the range of velocities in the only other detected nova super-remnant.

\section*{Acknowledgements}
MMS and JTG acknowledge the support of NSF award 2108234. KML is supported by the National Science Foundation under grants 1910001, 2107954, and 2108234. AK and LT acknowledge support from the National Research Foundation (NRF) of South Africa. AK acknowledges the Ministry of Science and Higher Education of the Russian Federation grant 075-15-2022-262 (13.MNPMU.21.0003). JCS and LS thank the European Southern Observatory for supporting their stay at the American Museum of Natural History via the SSDF program. MWH-K acknowledges a PDRA position funded by the UK Science and Technology Facilities Council (STFC). MJD and MWH-K receive funding from STFC grant number ST/S505559/1. MMS thanks Brad Schaefer for an early copy of his group's KT Eri paper, and helpful suggestions from Lisa Crause regarding SALT. We thank the Director of SALT for time allocation via the Director's Discretionary SALT program 2022-2-DDT-002. This work has made use of data from the European Space Agency (ESA) mission {\it Gaia} (\url{https://www.cosmos.esa.int/gaia}), processed by the {\it Gaia} Data Processing and Analysis Consortium (DPAC,
\url{https://www.cosmos.esa.int/web/gaia/dpac/consortium}). Funding for the DPAC has been provided by national institutions, in particular the institutions
participating in the {\it Gaia} Multilateral Agreement. This research made use of APLpy, an open-source plotting package for Python \citep{aplpy2012, aplpy2019}.

Facilities: Condor, SALT

Software: \textsc{APLpy} \citep{aplpy2012, aplpy2019}, \textsc{astropy} \citep{astropy:2013, astropy:2018, astropy:2022}, \textsc{gala} \citep{Gala2017, Price-Whelan2020}, \textsc{ndimage} \citep{Gomers2022}, \textsc{numpy} \citep{harris2020array}, \textsc{scikit-image} \citep{Vanderwalt2014}, \textsc{SFFT} \citep{Hu2022}

%%%%%%%%%%%%%%%%%%%%%%%%%%%%%%%%%%%%%%%%%%%%%%%%%%
\section*{Data Availability}
All data described here are available on the Condor web site https://condorarraytelescope.org or by contacting the corresponding author.

\newpage    



\bibliographystyle{mnras}
\bibliography{KTEri} 
% if your bibtex file is called example.bib

%\begin{thebibliography}{99}
%\end{thebibliography}

%%%%%%%%%%%%%%%%%%%%%%%%%%%%%%%%%%%%%%%%%%%%%%%%%%

%%%%%%%%%%%%%%%%% APPENDICES %%%%%%%%%%%%%%%%%%%%%

%\appendix


%%%%%%%%%%%%%%%%%%%%%%%%%%%%%%%%%%%%%%%%%%%%%%%%%%

% Don't change these lines
\bsp	% typesetting comment
\label{lastpage}
\end{document}